\documentclass[preprint, 1p, review,12pt]{elsarticle}
\usepackage{graphicx, color, url}
\usepackage{textcomp, wasysym,lineno}

\begin{document}

\begin{frontmatter}
\title{Energy-resolved fast neutron resonance radiography at CSNS}
\author[ihep,csns]{Zhixin Tan\corref{cor1}}
\ead{tanzhixin@ihep.ac.cn}
\author[ihep,csns]{Jingyu Tang}
\author[ihep,csns,skiprse]{Hantao Jing}
\author[ihep,csns]{Ruirui Fan}
\author[ihep,csns]{Qiang Li}
\author[ihep,csns]{Changjun Ning}
\author[ciae]{Jie Bao}
\author[ciae]{Xichao Ruan}
\author[ciae]{Guangyuan Luan}
\author[ustc]{Changqin Feng}
\author[nint]{Xianpeng Zhang}
\cortext[cor1]{Corresponding author}
\address[ihep]{Institute of High Energy Physics, Chinese Academy of Sciences (CAS), Beijing 100049, China}
\address[csns]{Dongguan Neutron Science Center, Dongguan 523803, China}
\address[ciae]{China Institute of Atomic Energy, Beijing, 102413, China}
\address[nint]{Northwest Institute of Nuclear Technology, Xi'an 710024, China}
\address[ustc]{University of Science and Technology of China, Hefei 230026, China}
\address[skiprse]{State Key Laboratory of Intense Pulsed Radiation Simulation and Effect (Northwest Institute of Nuclear Technology), Xi'an, 710024, China}

\begin{abstract}  
	The white neutron beamline at the China Spallation Neutron Source will be used mainly for nuclear data measurements. It will be characterized by high flux and broad energy spectra. To exploit the beamline as a neutron imaging source, we propose a liquid scintillator fiber array for fast neutron resonance radiography. The fiber detector unit has a small exposed area, which will limit the event counts and separate the events in time, thus satisfying the requirements for single-event time-of-flight (SEToF) measurement. The current study addresses the physical design criteria for ToF measurement, including flux estimation and detector response. Future development and potential application of the technology are also discussed.  
\end{abstract}

\begin{keyword}
Energy-resolved neutron imaging \sep Fast neutron resonance radiography \sep Time-of-Flight measurement \sep White Neutron Source\sep China Spallation Neutron Source \sep Non-destructive analysis
\end{keyword}

\end{frontmatter}

\section{Introduction}
\label{sec:intro}

\begin{figure}[hbt]
  \centering
  \includegraphics[width=0.85\textwidth]{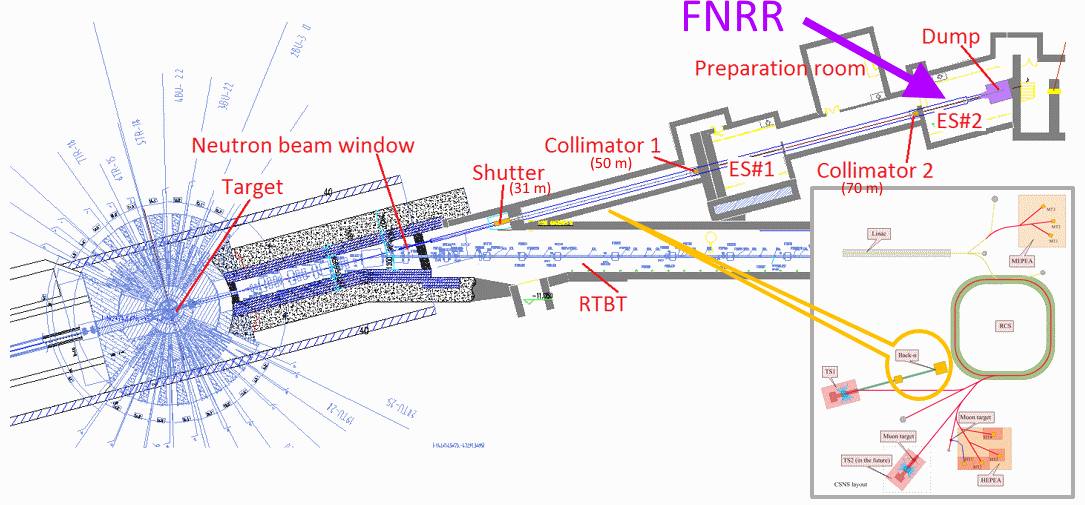}
  \caption{Location of the radiography facility at the CSNS}
  \label{fig:loc}
\end{figure}

The China Spallation Neutron Source (CSNS) project has started its  beam commissioning and is expected to put into service in spring 2018 ~\cite{Chen2016,Wei2009,WANGFW2013}. It will be among the most powerful spallation neutron sources in the world, with 100 kW in Phase one and 500 kW in Phase two. High flux neutrons is produced by a proton beam of 1.6 GeV and 25 Hz in repetition rate impinging a thick tungsten target. The China Spallation Neutron Source will serve both multi-disciplinary research based on neutron scattering techniques and nuclear data measurements based on white neutrons.  White neutron applications use the back-streaming neutrons along with the proton beamline that have a very wide energy spectrum from eV to hundreds of MeV~\cite{Jing2010,an2017,epj2017}. A beamline facility called ``Back-n WNS''  is being constructed to exploit the applications of back-streaming neutrons, as shown in Fig.~\ref{fig:loc}. Neutron resonance radiography is also found very suitable at Back-n in addition to nuclear data measurements.

Compared with X-ray technology, neutron radiography is often constrained by the source\cite{Strobl2009}, since the low-flux neutron sources will limit the count rate and lead to a low-quality image output. 
In our condition this problem is being addressed as the Back-n beamline will provide high flux neutrons. However, the white neutron beamline will also be distinguished by its wide spectra. In general, broad spectral sources are prone to artifacts that blur the image~\cite{Wang2016}. In that case, the white neutron beamline would be a `bad' imaging source as neutron source without majority component may result in a fuzzy imaging output. 
However, if we could tag the neutron transmission information with the corresponding particle energy via time-of-flight (ToF) measurement and obtain a continuous series of transmission images, more sample information could be collected in a single shot. This idea is similar to that of fast neutron resonance radiography (FNRR), proposed by Vartsky et al., where neutron snapshots are taken by a time-windowed device\cite{Vartsky2006,Vartsky2005,Mor2009,Dangendorf2009,Vartsky2010,Mor2012}.

In its proposed form, FNRR utilizes the total neutron cross-sections of elements (such as oxygen, carbon and nitrogen) in the fast-neutron spectral range for elemental analysis\cite{Vartsky2005,Chen2002}, which makes it of great value for non-destructive analysis of organic materials. Enlighted by this approach, we herein consider an imaging technique that exploits the fast-neutron component in our white neutron beamline. Firstly we focus on the neutron energy range from 0.5 ~MeV to 10~MeV. This energy range is the majority of our beamline and have not been touched by other neutron spallation source for energy-resolved neutron imaging.  Different from Vartsky's reports\cite{Vartsky2005},  our goal is more general: we directly measure the flight time of each neutron events and then re-constructing the neutron radiography images for energy-resolved neutron applications.
Certainly, position sensitive neutron transmission measurement with spectral resolution  is now  common  in cold and thermal neutron, where a great success have been achieved in the application of texture analysis, strain distribution, and  deformation process tracking~\cite{STR12201,SANTISTEBAN2012}.   Moreover,  a prototype of epithermal neutron resonance transmission  imaging  for archaeological researches had been demonstrated at the ISIS neutron source\cite{epithermal2009}. 
However,  for fast neutron, as far as we known there is no report on FNRR at spallation sources. 
Part of the reason is that the harsh requirement in time response make FNRR a challenging task in measurements.
Furthermore, the discrepancy in spectrum leads to a quite difference in the detector and signal processing, as well as their target application.

Our back-n beamline provides opportunities to develop new methods to exploit high-flux neutron radiography. In this paper, we describe an energy-resolved fast neutron resonance radiography model, which uses a liquid scintillator fiber detector array as the neutron imaging panel. The design of fiber neutron probes is inspired by the liquid fiber array in the inertial confinement fusion (ICF) for neutron snapshots and the time-resolved integrative optical neutron (TRION) system, as well as PSD experiments~\cite{Vartsky2005, Caillaud2012, Jamili2015}.

\section{Time-of-flight measurement and the model for neutron radiography}

Time-of-flight measurements, as the name implies, obtain the kinetic energy by timing the particle as it travels along a fixed distance~\cite{Coceva2002}. Neutrons with different energies will arrive at the detector, and thus trigger events, at different moments.
Nevertheless, an important assumption underlying the measurement is that the number of neutrons/events can be counted. Under ideal conditions, only a single particle event is processed within a given time interval, a scenario referred to as single-event ToF (SEToF) in Ref.~\cite{Vartsky2006}.  

\begin{figure}[thb]
  \centering
  \includegraphics[width=0.85\textwidth]{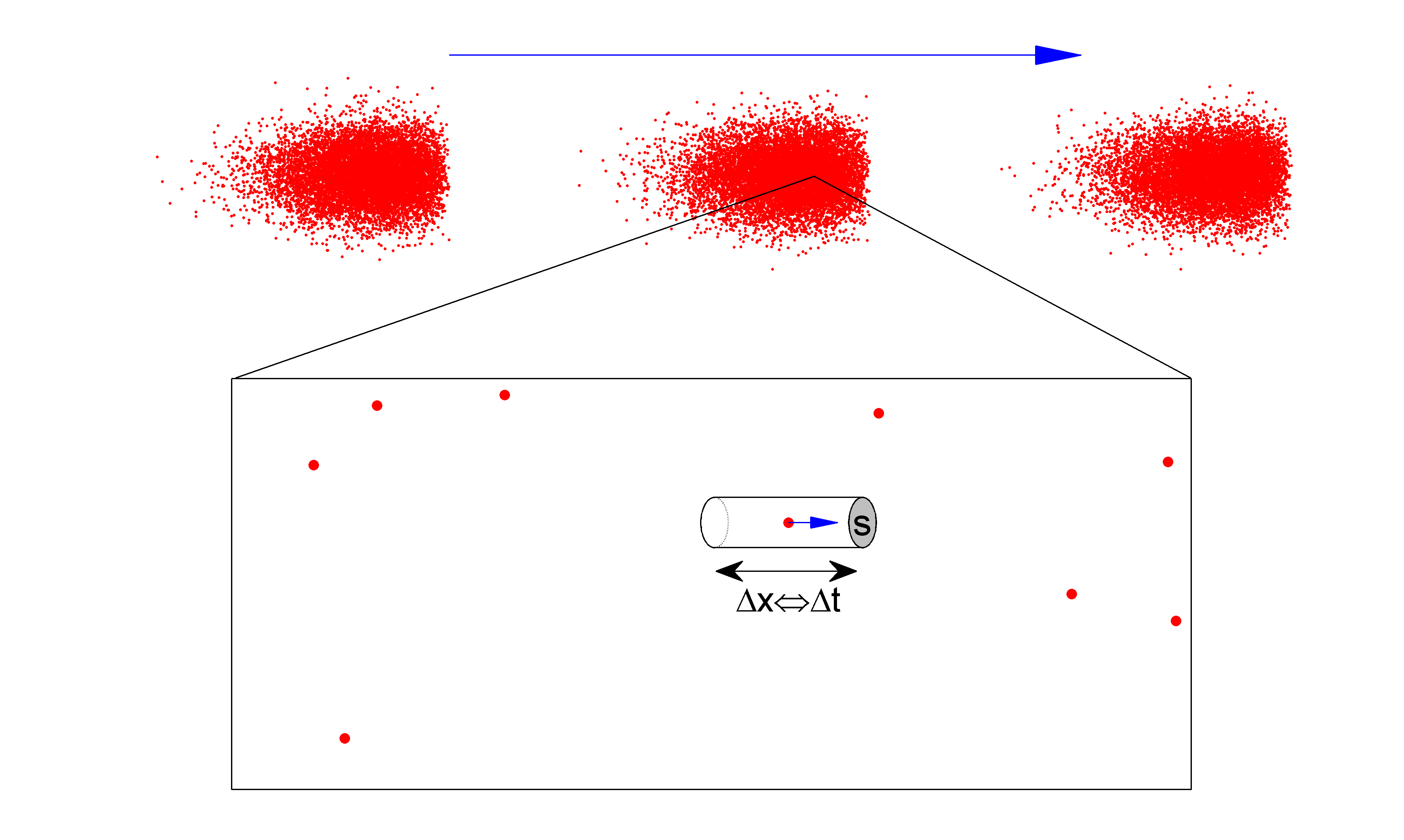}
  \caption{A schematic drawing of the pulsed neutron source and the logical image of the detection cell for SEToF.}
  \label{fig:rain}
\end{figure}

Fig.~\ref{fig:rain} presents a schematic drawing of a ToF measurement under a pulsed neutron source. Neutron bunches emitted from the source (on the left) fly to the detector (on the right). To be countable, the neutron events must be separable by the detector and its electronic readout. Each particle occupies a single cell, and the cells are separated in space -- or in time. Thus, for a ToF measurement, the detector cross-section and the corresponding time interval are critical for event counting, as illustrated in Fig.~\ref{fig:rain}. The challenge facing FNRR at the spallation source comes from the high neutron flux, which will exceed the capacity of conventional detectors since  a number of events will be generated within a time interval of several $\mu s$.
In this sense, the detection model of ToF is analogous to the counting of rain droplets, the difficulty of which varies under different circumstances. For example, it is easy to count droplets in a bucket under drizzle, whereas it is difficult to count droplets in a pond during the rainstorm.  

A SEToF measurement strictly requires less than one event per cell in the electronic cycle of the detector. To achieve this, a small detector cross-section with a fast time response is required for a SEToF measurement under a high-flux neutron source. 
Since the spallation reaction is a random process and the flux represents its average activity, our analysis proceeds in a statistically averaged manner. Thus, the average count in a single electronic interval should be much less than one, written as
\begin{equation}
  \centering
     \bar{N} = \eta \cdot \bar{F} \cdot S \cdot \Delta t \ll  1 
  \label{eq:average}
\end{equation}
where $\eta$ is the detector efficiency, $\bar{F}$ is the instantaneous neutron flux, $S$ is the detector cross-section and $\Delta t$ is the time interval of the electronic cycle. Together, the time interval and the detector cross-section constitute the detection cell.
For a ToF measurement, organic scintillators with a decay time of several nanosecond are used to ensure a fast response, so the time interval is defined by the scintillator detector and its electronic readout. Therefore, the detector cross-section is the critical parameter to limit the number of event signals in the experimental design. In the following section, we present our proposed configuration for FNRR and discuss the parameters in Eq.~\ref{eq:average}.

\section{Configuration for fast neutron resonance radiography}

The neutron imaging facility will be built at Endstation-2\# along the white neutron beamline, as indicated by the solid purple arrow in Fig.\ref{fig:loc}. The neutron flight path will be 77.5 m in length.
To reach the detector, neutrons of energy 1 ~MeV will take 5600 ~ns, while those of 10 ~MeV will take 1786 ~ns. In addition, the spallation reaction will provide a natural start time $T_0$ for ToF measurement because a gamma signal  will be produced simultaneously during spallation reactions. Traveling at light speed, this gamma radiation will reach the target 258~ns after the burst, well ahead from the neutron scintillation  signals. This will allow efficient and precise ToF measurements for all neutrons. 

\begin{figure*}[bth]
  \centering
  \includegraphics[width=0.95\textwidth]{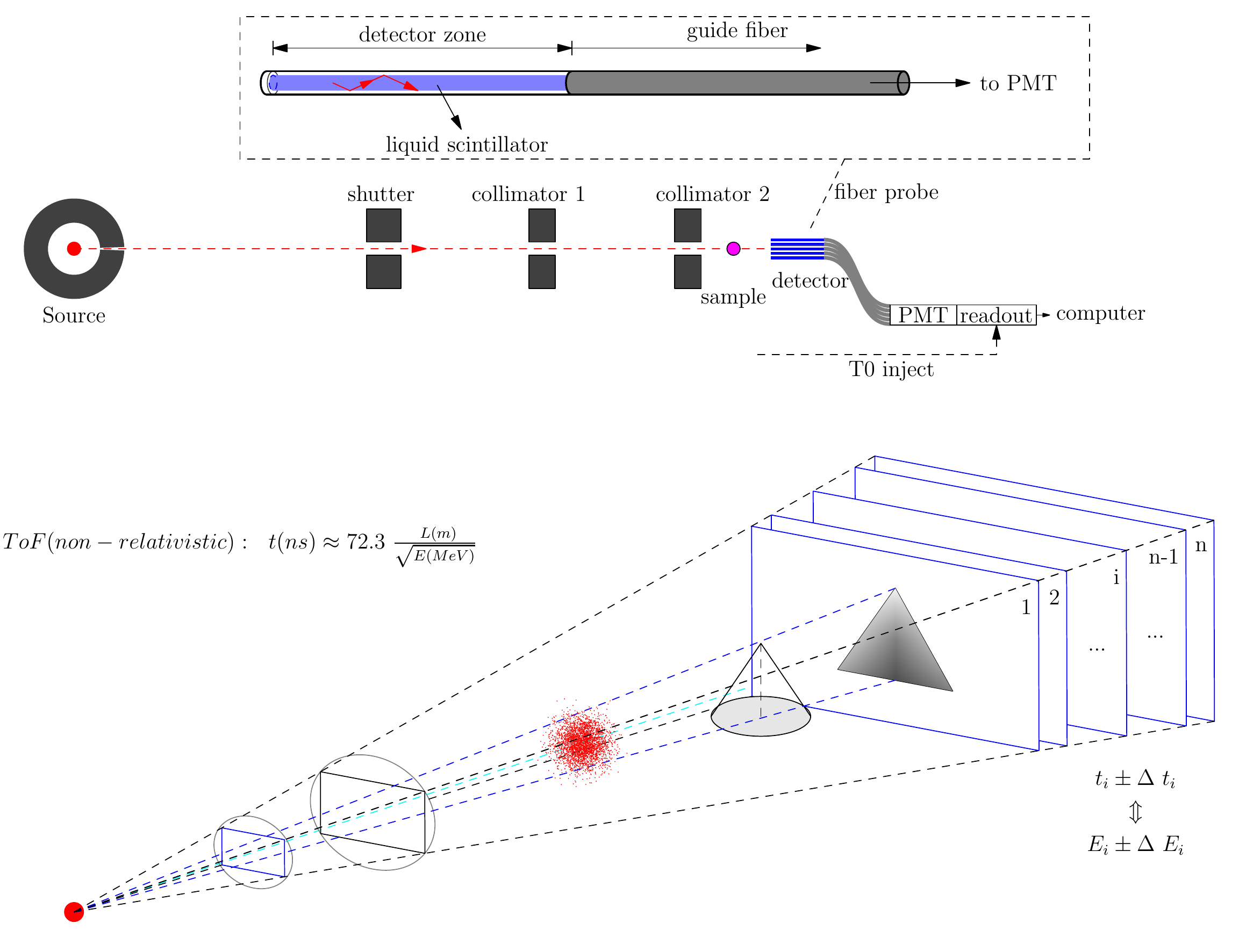}
  \caption{Schematic drawing of fast neutron resonance radiography at the CSNS and its prospective results}
  \label{fig:scheme}
\end{figure*}

Figure~\ref{fig:scheme} presents a schematic drawing of the proposed FNRR at the Back-n. Our system consists of several components: the neutron beam, sample,  scintillator probe, multi-anode photo-amplifier (MAPMT), electronic readout, and data acquisition. The neutron sensitive  probe is a liquid scintillator fiber array with 256 units. One unit of the scintillator probe is zoomed out within the dashed line for illustration. The liquid scintillator in probe unit has a length of 40 mm, a small cross-section of  $0.5$ mm in diameter, which will effectively capture the neutrons without overlapping  in time-scale.  
During experiment, collimated neutrons penetrate through the sample and hit scintillators. Unaffected neutrons will have a good probability to be detected. Photons stimulated during the scintillation process will be diverted into the PMT by guide fibers.  The multi-anode PMT will transform the optical signals into electrical signals and a gain of $10^6$ is expected. The readout electronics will manage electrical signals and extract the time information.

While the sample is irradiated by the pulsed beam, event counts will be accumulated and their flight times recorded. Neutron transmission spectra can then be derived from the flight times before and after sample irradiation. The prospective result will be a series of energy-resolved neutron transmission images, sorted in energy sections, as illustrated at the bottom of Fig.~\ref{fig:scheme}. Because each pixel will come from an individual fiber detector, no image diffusion will take place. Every pixel of the image will come from the sum total of transmitted neutrons, which will depend on the total cross-section of the materials encountered during the neutron flight path. Thus, the change in count caused by the in-situ sample can be measured, and consequently post-processing for elemental analysis can be carried out.  

\begin{figure}[!ht]
  \centering
  \includegraphics[height=0.81\textheight]{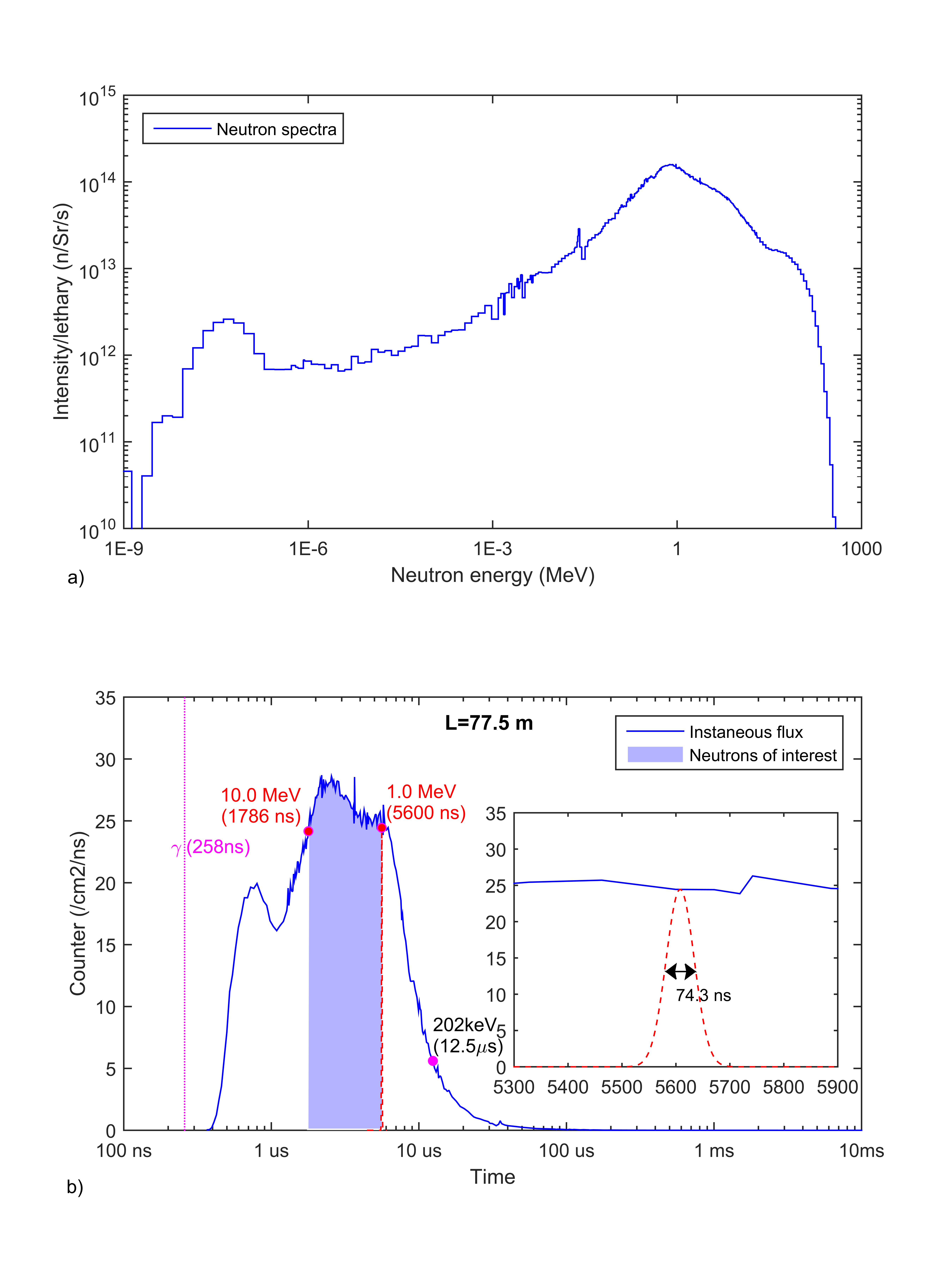}  
  \caption{a) Spectrum of the white neutron beamline; b) Instantaneous neutron flux at Endstation-2\#}
  \label{fig:tof}
\end{figure}

The beamline flux is estimated as $6.8 \times 10^6$ n/cm$^2$/s at the sample position with a proton beam power of 50 kW in single-bunch mode~\cite{an2017,epj2017}.  
{The flux can be derived from the CSNS neutron yield.  The basic parameter of CSNS are listed in Citation ~\cite{WANGFW2013} table 1. The average proton current is 62.5~$\mu A$ and approximate $1.63\times 10^{13}$ proton per bunch will be delivered to the target. 
The expected neutron yield is  $\Phi \approx 1.63\times10^{13} \times (2\times25)\times 25 = 2\times10^{16} $~n/s, since the CSNS runs at 25Hz, each pulse contains two bunches, and approximate 25 neutron produced from a high-energy proton hit. 
The back-n beamline is a direct view to the target core and there is no moderator at the beamline head. So a coarse flux estimate at 80~m distance will be $\Phi/(4 \pi R^2) = 2.5\times 10^7~  \mathrm{n/cm^2/s }$, and  $1.25\times 10^7~  \mathrm{n/cm^2/s }$ under single-bunch mode.  Apparently, our nominal flux ($6.8\times 10^6~  \mathrm{n/cm^2/s }$)  is much less than the coarse result, whereas the collimation should account for the $\sim50\%$ loss.}
As shown in Fig.\ref{fig:loc}, the beamline is collimated by a shutter and two collimators at a distance of 31~m, 50~m and 70~m, respectively. The collimator apertures are 50~mm and 58~mm, while beam spots of 50 mm and 60~mm are expected\cite{epj2017}. Therefore the beam divergence is only 0.29 degree.
In short, the back-n beamline is a bright neutron beamline with small beam divergence, which enables high-quality neutron imaging. 

The predicted spectrum of back-n beamline is shown in Fig.~\ref{fig:tof} a). It should be noted that the x- and y-axes are in log scale. The most common neutron energy is approximately 1.0~MeV, and fast neutrons is the main component.
Based on the back-n spectra and beamline flux, we can derive its instantaneous  flux, as shown in Fig.~\ref{fig:tof} b). The x-axis is arrival time in log scale, and the average neutron flux within a square centimeter per nanosecond are presented with its corresponding time.
The purple vertical line marks the time at which the gamma signal is recorded. Several specific energy points are labeled with their flight time in brackets.
In Fig.~\ref{fig:tof} b), it can be qualitatively seen that the beamline is densely structured, in the sense that fast neutrons are not only plentiful but also arrive within a short interval of several microseconds. The flux reaches a maximum value of  28.7~neutrons/cm$^2$/ns at 2.194~$\mu s$, decreasing steadily before the most popular energy 1~MeV,  then drops rapidly from 5.6~$\mu s$ to 12.5~$\mu s$, corresponding to neutron energies of 1~MeV and 202~keV. As shown, the flux is about 25~neutrons/cm$^2$/ns in the range from 10~MeV to 1~MeV. A rough estimate may help our understanding.
Assuming $6.8 \times 10^6$ neutrons per second, this suggests that there are over $2.72\times 10^5$ neutrons within single bunch. Among them, neutron with energies ranging from 1~MeV to 10~MeV accounts for about one third, i.e., approximately $9.07\times 10^4$ neutrons. Due to the use of a pulsed neutron source, these neutrons are burst from the same pulse and subsequently diverge during their flight. The corresponding time interval is $\Delta T = 5600 - 1786 = 3814$~ns. So the instantaneous neutron flux is about 23.8~n/cm$^2$/ns. This estimate confirms our result that a dense frontier will lead the neutron bunch.

The FNRR measurement requires not only a sufficiently high average event count within a short interval for good statistics, but also a fast response detector to precisely determine the particle arrival time. Most commonly, ToF measurements use organic scintillators, which are often plastic or liquid. The candidate liquid scintillator is EJ-301, whose mean decay times of first 3 components are 3.16~ns, 32.3~ns and 270
~ns respectively~\cite{ej301}. EJ-301 has a high scintillation efficiency (12000 photons/1 MeV e$^-$) and high light output (78\% of the anthracene). The performance of EJ-301 is suggest to be identical to the widely reported NE-213 and exhibits all of the properties of the latter. 
In the following text, we do not distinguish between these two liquid scintillators, and treat them as the same material.
EJ-301 is been popular used for neutron detection as it exhibits pulse shape discrimination (PSD) properties. The PSD property preserves the possibility of extracting fast neutron signals under a background of gamma radiation. In Eq.~\ref{eq:average}, two parameters, the detector efficiency and the time interval, are determined by the physical detector. To simplify our investigation, we use previously determined results to estimate these parameters. 

The time interval $\Delta t$ is used to distinguish particle events, for which the original time response record of the detector is needed. Fortunately, in the past few years many studies have done on the subject of PSD and the time response of detectors were measured in detail.
R.F. Lang et al. reported their improvement on PSD study with EJ-301~\cite{Lang2017}. After amplified by the PMT, scintillation signals were digitized by CAEN DT5751, which records the response curve at 1 GHz with a resolution of 10 bits.  Figure 3 in citation~\cite{Lang2017}  present the response to neutron and the fast portion of a neutron event is suggested to be about 20~ns in width.
Similar result is also shown in Figure 2 in citation~\cite{CESTER2014} . Another neutron response curve  with clear time ticks was presented  by Jamili et al~\cite{Jamili2015}.  The liquid scintillator used  in experiments is NE-213, which is suggest to be identical to  EJ-301.  The complete curve response was record by high-speed oscilloscope in Figure 3 in citation ~\cite{Jamili2015} and the fast portion is 20~ns  in width.
 Apparently, response curves in these two reports are very similar in shape and their time ticks. Since EJ-301 has multi- scintillation components with time overlap, the curves are suggested to be reasonable. It shows a fast rise due to the rapid accumulation of photoelectrons, and a quick falloff as the discharge is launched. A slow depletion follows after most of the electrons have been ejected. 
For ToF measurements, the left-hand edge of the curve marks the arrival of a neutron event. Therefore  a time interval of $\Delta t = 20$~ns is suggested to be a well-grounded value for the cell length, as the peak separation of two events is expected to be less than that.

Liquid scintillator have a reputation for its high detection efficiency and widely used in neutron detections.  The efficiency depends on the incident neutron energy and the threshold setting. Neutron  kinetic energy has two affects on  the detection efficiency. 
On the one hand, as the incident energy grows, more ionization photons will be produced, thus increasing the detection efficiency. On the other hand,  with the increase of energy, the faster the neutron velocity, the smaller the cross-section, the lower the detection efficiency. 
These two effects get a balance at the spectrum of several MeV,  where energetic neutron  have a maximum possibility to be detected.  As shown in Figure 4 in citation ~\cite{DROSG1972},  the detection efficiency is as high as 30 \% at 256 keV threshold, and the event energy is up to 2~MeV~\cite{Cecil1979}.
 Then the efficiency reduces gradually to a plateau  below 20\% as the neutron energy increases from 2 MeV to 26 MeV. 
It also shows that the threshold value greatly affect the efficiency since  curves decrease correspondingly at higher thresholds.
Many years later, with an advance in nuclear electronics and measurement,  experiments on the lower threshold were done on the Colonna et al. in BC-501~\cite{Colonna1998}, another scintillator, which is suggested to be equivalent to EJ-301.
Fig.4 in citation~\cite{Colonna1998} shows similar curve shapes  and  efficiency up to 75\%  at a threshold of 10 keVee was recorded.  
Certainly, these reports with clear efficiency data are done on large-size liquid scintillators. But they demonstrate liquid scintillator has a good efficiency on event counting.  
An related FNRR experiment with tiny liquid capillary is reported by Vartsky team in 2012~\cite{Mor2012}.  
The capillary device is 3 cm long with inner average diameter 11 $\mu m$ and  its triangular corners were excluded from the filling. Liquid scintillator used here is EJ-309, which servers as an alternate to EJ-301 for similar optical performance, but with low chemical toxicity and high flash point.  The neutron detection efficiency was estimated as 10-20\% in the conclusion paragraph in citation ~\cite{Mor2012}.  
For the sake of conservatism, a detection efficiency of 10\%  is chosen for our calculation while details are omitted.

The inner diameter of the detector is chosen as 500~$\mu m$, corresponding to a  narrow capillary with a liquid scintillator filling.  Thus, the average count within a cell in Eq.~\ref{eq:ave} is now
\begin{eqnarray}
  \centering
  \bar{N} &<& \eta \cdot \bar{F} \cdot S \cdot \Delta t                                                           \nonumber   \\
      &\approx& 10\% \times 23.8 \times (\frac{\pi}{4} \times 0.05^2) \times 20       \nonumber \\
      &=&  0.0935  ~~~~~~~~~~\ll  1 
  \label{eq:ave}
\end{eqnarray}
With  this small detector cross-section, the averaged count is much less than 1. Clearly, this is beneficial for the ToF measurement, as the events are widely separated in time.
Of course, for statistical validity, the event count rate is also important in neutron imaging. In current configuration, the event count of fast neutrons within a fiber per pulse will be around $  N_c =  23.8  \times  \eta \cdot S \cdot \Delta T = 17.8 $. The time duration $\Delta T$ is used to estimate the event count in experimental consideration.
Another benefit is that the dependence of time interval $\Delta t$ is removed in this estimate. Therefore, each pixel will make over 446 counts per second, and $2.67\times 10^5$ events will be recorded after ten minutes exposure.

\section{The probability of signal overlap}

The above section presented a description of our model for a single event ToF measurement. However, some important issues remain unresolved. As the name implies, a SEToF measurement requires less than one signal per cell. Since the spallation source emits neutrons in a random manner, there is always a possibility that two signals will register within one cell. So the key problem for SEToF is minimizing the probability of event overlap, and particularly, the probability of two events signals overlapping. Therefore, a proper estimate of the probability of signal overlap is the first step.

The probability estimation is analogous to the process of placing balls into boxes. For example, suppose that we place 2 balls into $N$ boxes, and then calculate the probability that these 2 balls are both placed in the same box, which is  among the first {\emph{\textbf m}} boxes. Thus, the probability is written as $p = m\cdot  {\mathrm{p}^1_\textsc{\tiny N}} \cdot  {\mathrm{p}^1_\textsc{\tiny N}}$.

An important concept here is the matching length, which is equivalent to the number of boxes \emph{\textbf m} in the example above. In our case the matching length is the time duration in which particle events of similar energy may take place in the same cell. For a neutron of fixed energy, it concerns about what energy does another neutron own and how long do these `similar' neutrons span in time-scale. The matching length is estimated as follows. First, we choose a spectral range at a specific flight time within a timespan of double cell length (40~ns) to include both forward and backward overlapping from a fixed point. In specific, particles with a flight time of 5600 ns have an energy of 1~MeV$\pm$ 7.3~keV. Second, we estimate the time duration of these neutrons. The time duration depends on the time structure of the spallation source. Because spallation reaction is a cascade process, low-energy neutrons may be generated from high-energy particle impacts. Obviously, the time duration of the low-energy neutrons will be much longer than that of those with high energy.
This differs from the case of an accelerator-driven fission source, where neutrons of varying energy have same duration.  Jing et al had reported the time structure of the back-n beamline~\cite{Jing2010}, where neutrons with an energy of 0.9999-1.010~MeV is 52.1~ns in time scale. For simplicity, the time duration of the neutron envelope with energies of 1 MeV $\pm$ 7.2~keV is extrapolated as $ t_m=52.1*(14.4/10.1) = 74.3~\mathrm{ns}$, as illustrated in the inset of Fig. 4b).

Assuming that the average count is 0.1, we can expect that a particle event will be detected by the scintillator fiber within 10 intervals. It also means the probability of an event within any one signal interval is 10\%. Here, each signal interval corresponds to a box, and an event to a ball. Thus, the probability of two signals overlapping is
\begin{equation}
    p = m \cdot  (\frac{\mathrm{p}^1_\textsc{\tiny N}}{2} )^2 = (\frac{74.3}{20}) \cdot (\frac{9.35 \%} {2}) ^2 = 0.81 \%
  \label{eq:pos}
\end{equation}
The denominator 2 accounts for the fact that double interval lengths are needed for 2 events. Thus, the probability of signal overlap for particles with energies around 1~MeV is less than 1 \%.  Since 1~MeV neutrons are the most popular particles, with the longest matching lengths in our configuration, they have the greatest probability of overlap. Therefore, across the spectrum, the overlap probability will be less than 1\%, implying that less than 1 signal overlap will occur per hundred events recorded. 
With even smaller diameters, the probability would be reduced further. However, the smaller cross-section  will decrease the count rate, causing the problem of event discrimination  under background noise. Thus, the proper diameter is a compromise between the minimal event overlapping and the event count ($N_c$) per unit.

In short,  we  put approximate 18 events in a time duration of 3814~ns and their overlap possibility will  less than 1\%.  If these events are evenly distributed, each event takes about 212~ns, and one signal occupy a time length of 20~ns.  So the low overlap probability should be attributed to the very short match length, which is determined by  the excellent time structure of the beamline, as described by Jing. et al~\cite{Jing2010}. 
In our model, the most uncertain value is the efficiency, which depends heavily on the implement specifications. Furthermore, a lateral penetration of the recoiled proton may decrease the efficiency for a small cross-section. However, since forward-scattering protons is the majority and our efficiency is a conservative estimate based on the Vartsky's experiment,  a 10\% efficiency is sensible.  
At a lower efficiency of 5\%, our model can work even though the number of counts reduced to half.
Another variable parameter is the power of CSNS (or neutron flux),  which greatly affects our design, as shown in the above deduction. In fact, the CSNS will undergo a very long period of gradual power escalation from its beam commissioning to a full power operation. Therefore, 
although the detector has a measurement range, it also should adapt to the change of running power and maintain its stability within a certain stage.
As a proposal, our design is based on the probabilistic model, and these calculations are preliminary, especially in terms of detection efficiency. Although the quoted data are from reliable experimental reports, specific values should be adjusted according to the experimental requirements. 
Related technologies are being developed, and in this article we will focus on the physical model itself and try to provide a framework for FNRR development at the back-n beamline.

\section{Discussion}
In the following, we will discuss various aspects of the facility and its performance. In addition, we will compare it with TRION and make a prediction of its potential applicability.

\subsection{Fiber scintillator probe and electronic readout}

The fiber liquid scintillator array has been demonstrated in previous experiments~\cite{Caillaud2012,Mor2012}.
 Similarly, in our design the liquid scintillator is poured into a narrow silicon capillary, as depicted in the inset in Fig.~\ref{fig:scheme}. To avoid radiation damage, signal photons are guided through a multi-mode fiber, which will be spliced at one end of the capillary. To imitate a conventional liquid scintillator detector, the detector cell (including the fiber probe, the guide fiber and the other end of the capillary) are proposed to coat with an aluminum layer on their outside, which confines all signal photons and guide them to the MAPMT. In this way the fiber probe is equivalent to a minimized liquid scintillator detector.

Compared with a liquid fiber probe, a commercial plastic scintillator fiber would greatly simplify the detector manufacturing. However, in our condition the liquid scintillator implementation offers three advantages. 
First, the luminous efficiency of a liquid scintillator is significantly higher than that of a plastic fiber. Specifically, the liquid scintillator EJ-301 emits 12000 photons under 1~MeV $e^-$  radiation while BCF-10 plastic fiber emits only 8000 photons~\cite{ej301,bcf10}. This is especially important under small signal conditions. 
Second, liquid scintillator support various diameters while the dimensions of plastic fibers are limited by suppliers and their production technology.  Apparently, it is relatively economical to produce an experimental fiber probe with liquid scintillator in variable size options.  In specific,  silicon capillary supports a higher spatial resolution which is desirable with a even-high  flux for 500~kW in CSNS-II.
Third, for  future development, a Boron-loaded liquid scintillator EJ-339 will be demanded to cover full white spectra~\cite{PINO2014}. Therefore, fiber probe with liquid scintillator will be convenient for future upgrade.

\subsection{ Energy resolution and spatial resolution}

The critical feature of our radiography design is the energy-resolved imaging.
These images depend on the high speed electronic readout, which is capable of processing a fast response to scintillation in the time scale of the nano-seconds, allowing the arrival of neutron events to be accurately determined.
Because the time interval of the cell is only 20~ns, the variation of signal rising edge should be no more than half of the signal FWHM. Thus, the time resolution, which is defined by the detector and electronics, should be less than 5~ns, and a  time  resolution of 3~ns is expected in experiments. {Comparing with previous report\cite{CESTER2014}, our resolution estimate is conservative as the complexity of the electronic readout in hundreds of channels is considered.}  Based on the above presumption, the system energy resolution can be interpreted at each point. Thanks to the high flux and the long flight distance, the energy resolution will reach 1.8~keV at 1~MeV, and 56.9~keV at 10.0~MeV, with 5~ns temporal resolution. Obviously,  these results is much better than the requirement for FNRR instruments proposed by Vartsky~\cite{Vartsky2006}.

Another important feature of this design is the continuous slicing of the spectra. For resonance radiography, a dense array is crucial for elemental analysis. Thus the count rate and system efficiency should be  evaluated. Different from the TRIONs, which selectively shot several images at specific moments in high frequency, our configuration randomly samples events across the entire bunch. 
The temporal efficiency in our configuration seems low, because we have to limit the number of events and keep them separated. However,  the selective images from TRIONs does not make sure events captured at certain positions, while we use all events from across the time scale and no mechanical change. Thus our configuration is efficient in temporal accumulation.
Overall, although we have only 256 units now, we have the advantage of single shot count in time while TRIONs preserve their high spatial resolution. In this sense, we are moving in a different direction from the developers of TRION. In fact, we prioritize the temporal handling above the spatial resolution, as we consider that the latter can be improved in the future.

Thus, a discussion on spatial resolution is now appropriate. Certainly, the narrow cross-section creates a problem in imaging because the real detected area is limited. But the probe does not have to one close the other. In our plan, the fiber probe is placed in pitch to represent a spatial distribution. Each pixel corresponds to a cell. Thus, the spatial resolution is determined by the pitch. 
To cover a inner square  of $42\times42$ mm$^2$ within a 60 mm diameter spot, the spatial resolution will be about 2.8~ mm.  This performance is poor but decent for energy-resolved researches. Furthermore, it is  possible to get a higher resolution to 1.0~mm by arranging  the detector probes within  a smaller area. As a initial model with only 256 channels, we plan to develop neutron imaging  researches with small objects, such as battery, coins et al.  
Another common method to improve spatial resolution is  using a step motor to move fiber probes a small distance, which is feasible under a stable flux. The deficiency of current design is the limited number of probes and channels, which results in a large pitch value and a low-resolution imaging output.
In another perspective, the poor resolution also represents a great potential for development, as demonstrated by previous experiments~\cite{ Mor2012,Caillaud2012}.  
The spatial resolution is anticipated to be significantly improved in future by developments in the mass production of detectors and the parallelization of amplification channels. 
Theoretically, the configuration of  small cross-section scintillator probe, which is demanded by ToF measurements, and  the high flux spallation neutron source together constitute the solution for high resolution FNRR. For future application development, a densely packed array of fiber probes is expected to record high-resolution images directly and finally overcome this problem.  

\subsection{ Future development for  non-destructive analysis }
Continuous, dense, energy-resolved neutron transmission images, combined with computerized tomography technology, allow the transmission of three-dimensional information on a sample. The energy-resolved attenuation coefficient $\mu_{x,y,z}~(E)$ serves as a fingerprint of the sample nuclei at each grid point. Through comparison with the full-section data in the nuclear database, a three-dimensional non-destructive analysis  (NDA) of the elemental composition of the sample will be achieved. Thus the interaction of neutron with the nucleus gives a direct evidence of the material composition. This marks an essential difference from the conventional density imaging from X-ray method, which is an attenuation of x-ray through the electron clouds outside the nucleus. Obviously, this neutron NDA technique will be direct and more powerful than X-ray method. 
With a solid theoretical foundation, advances in experiments  are expected to expand the potential applications of this facility into many research fields, such as isotope sample analysis,  archaelogy and other applications. 
The spectral difference from FP5@LANSCE may help our FNRR to be a unique place for organic material researches. In the future, Boron10-loaded/Lithium6-loaded liquid scintillator will be implemented as neutron probes to cover whole spectra of white neutron beamline~\cite{PINO2014,Bass2013}, raising the possibility for new discoveries.

\section{Conclusion}

In conclusion, a facility based on a liquid scintillator fiber array was proposed for fast neutron resonance radiography at CSNS. By determining the requirements of single-event ToF measurement and the characteristics of neutron beamline, the detector dimensions were designed to exclude the overlapping of events as far as possible. The facility’s performance was evaluated, including the energy and spatial resolution. As a new facility with continuous, dense, energy-resolved neutron transmission imaging output,  the back-n FNRR will have promising application prospects in many fields.

\section{Acknowledgement}
The authors would like to acknowledge the support from the CSNS Engineering Project and National Key Research Program of China (Grant No. 2016YFA0401601). 

\bibliography{slc3}
\bibliographystyle{elsarticle-num} 

\end{document}